# Hidden walls: STEM course barriers identified by students with disabilities

Westley James, Kamryn Lamons, Roberto Spilka, Caroline Bustamante, Erin Scanlon and Jacquelyn J. Chini

*Physics Department, University of Central Florida, Orlando, FL, 32816*

Historically, non-disabled individuals have viewed disability as a personal deficit requiring change to the "disabled" individual. However, models have emerged from disability activists and disabled intellectuals that emphasize the role of disabling social structures in preventing or hindering equal access across the ability continuum. We used the social relational proposition, which situates disability within the interaction of impairments and particular social structures, to identify disabling structures in introductory STEM courses. We conducted interviews with nine students who identified with a range of impairments about their experiences in introductory STEM courses. We assembled a diverse research team and analyzed the interviews through phenomenological analysis. Participants reported course barriers that prevented effective engagement with course content. These barriers resulted in challenges with time management as well as feelings of stress and anxiety. We discuss recommendations for supporting students to more effectively engage with introductory STEM courses.

## I. INTRODUCTION

Students with disabilities make up between 10-20%[1] of the postsecondary population [1,2], and course barriers exist which prevent this population from equal access to learning. For example, Mullins and Preyde interviewed postsecondary students with disabilities and identified barriers such as negative faculty attitudes towards disability and accommodation use and courses not being designed with students with disabilities in mind [3]. Little work has investigated the barriers students with disabilities experience specifically in STEM courses [4]. By identifying specific barriers present in STEM courses, more focused recommendations can be made regarding how to better support students with disabilities.

Different models of disability promote different societal actions and different research agendas [5]. This study is underpinned with the social relational perspective [6], which posits "disability to be those restriction of activity that result from the exercise of power to exclude: disability only comes into being when restrictions of activity are wholly social in origin" (p.29) [7]. Thomas separates socially constructed limitations (i.e., disability) from restrictions of activity that arise directly from impairment, which she terms "impairment effects". We diverge slightly from our understanding of Thomas's perspective to acknowledge the positive characteristics (in addition to limitations) that our participants associated with their disability diagnosis. Thus, we use the term "diagnosis characteristics" to include personal characteristics participants associated with their diagnosis, and to distinguish from the negative connotation associated with the word "impairment".

Thomas's social relational perspective highlights the importance of disentangling socially constructed disability from impairment effects (in our model, diagnosis characteristics). We use this perspective to emphasize instructors' power to reduce forms of social oppression that inhibit full participation in the physics community by individuals with impairments. For example, an individual with a mobility impairment may be unable to physically climb stairs, but this limitation is only "disabling" if a building lacks an elevator the individual can use. Thus, access can be provided by changing the building without changing the individual.

In this paper we will report barriers in STEM courses for students with disabilities and conclude by providing recommendations for how instructors can begin identifying and addressing these barriers by using a framework called Universal Design for Learning [8].

## II. METHODS AND ANALYSIS

We recruited participants enrolled in introductory physics or chemistry courses at a large southeastern research-intensive university to participate in two, one-hour semi-structured interviews conducted at the beginning and the end of the semester they took the course (though not all participants included in our analysis completed both interviews). Some participants were in SCALE-UP physics courses, a "flipped" course format, often held in a specially designed room, where emphasis is placed on interactive team learning [9]. In the interviews, we asked questions about the participants' experiences as a student with disabilities in postsecondary STEM courses.

Participants were recruited via emails sent by the disability services office and by faculty teaching the courses. This allowed both students who had registered to use accommodations in their courses and those who were not registered to participate in the study. Thirty-one participants who identified with a disability diagnosis were recruited; we selected nine participants' transcripts for analysis because these participants made explicit statements about their diagnosis characteristics and they talked about the interaction between their diagnosis characteristics and their courses. The participants collectively represented a variety of disability diagnoses and were diagnosed at different ages, as presented in Table 1. We chose to analyze the interviews through interpretative phenomenological analysis, which suggested focusing on smaller sample sizes, averaging around six participants, to promote depth versus generalization of findings. Our sample extends to nine participants to account for the wide variation in disability diagnosis [10]. Limitations from our recruitment are that not every existing disability diagnosis is represented among our participants, not every disability diagnosis is represented equally, and all participants came from the same university.

The lead researcher (W.J.) conducted all of the interviews and identifies as a white, non-disabled male. Four of the authors (W.J., K.L., R.S., and C.B.) cooperatively conducted the analysis. We analyzed each transcript through interpretive phenomenological analysis (IPA) [10] via the following steps: 1) the researchers read through the transcript to familiarize themselves with the data; 2) researchers identified statements where participants talked about their diagnosis characteristics or how their diagnosis characteristics interacted with their college experience generating comments; 3) researchers generated succinct phrasing for each comment which highlighted the main points (referred to as themes); 4) W.J. selected themes which included the words "chem", "chemistry", "physics", or "STEM"; 5) researchers identified the barriers and supports most commonly represented by each participant and their selected themes; and 6) researchers identified the most common barriers and supports expressed across all participants. To acknowledge the unique perspective and experiences of the participants, steps 1-5 were conducted independently for each participant and transcripts were assumed to be independent (i.e., findings from the first transcript analyzed were not carried forward to the next transcript during analysis). Step 1 was conducted independently by each researcher, but every other step was conducted in group meetings where the graduate student (W.J.) and two of the undergraduate researchers (either K.L. and R.S. or K.L. and C.B.) were present.

TABLE 1. Demographics for participants (pt) (identifier, diagnosis, and age diagnosed)

| Pt | Diagnosis shared by pt | Age diagnosed (years) |
|----|------------------------|-----------------------|
| 1 | Attention-deficit/hyperactivity disorder (ADHD) | 16 |
| 2 | Adult ADHD | 22 |
| 3 | ADHD | 11 |
| 4 | Bipolar disorder, anxiety, clinical depression | 12 |
| 5 | X-Linked juvenile retinoschisis | 1 |
| 6 | Asperger's, depression, migraines, sleep apnea, driving anxiety | 12 |
| 7 | Dysthymia and generalized anxiety | 15 |
| 8 | Autism | NA |
| 9 | Severe anxiety | NA (during college) |

---

[1] The statistics for students with disabilities' college enrollment varies across studies due to their methods of data collection and definitions of diagnosis categories.

The researchers represented a spectrum of disability diagnosis, including non-disabled, diagnosed, and undiagnosed but identifying with diagnosis characteristics. We also represent both graduate and undergraduate students and come from physics and non-physics STEM majors. A recommended practice of the IPA process is for researchers to avoid taking their assumptions into the analysis (i.e., bracketing) as these will interfere with interpreting how participants interpret their experiences [11]. By having a diversity of researchers co-generating interpretations, researchers are more likely to be confronted with differing interpretations which require each researcher to reflect and defend how the participant's words are the basis of the researcher's interpretation [12]. When using IPA, we also interpret participants' statements by asking questions of the transcripts such as "Do I have a sense of something going on here that maybe the participants themselves are less aware of?" [10]. An example in this study's context is that a participant may be unaware of the perspective that disability is located in societal barriers, but when analyzing their transcript, we find interactions between diagnosis characteristics and course practices identified as harmful.

To prepare for analysis, the graduate student trained the undergraduate researchers about IPA by co-generating comments and themes for a transcript not included in this study. In acknowledgement of the power dynamics[2] present, norms were established and repeated by the lead researcher throughout the analysis process that every researcher's interpretations were of equal merit. Researchers were encouraged to express and defend their interpretations, and the research team discussed disagreements until consensus among the three researchers was met. By having to express and defend our interpretations with others involved in the analysis process, we perform a form of ongoing peer review, a practice which Creswell and Poth identify as supportive in building trustworthy interpretations [13].

## III. FINDINGS

From our analysis, we found that participants commonly linked their diagnosis characteristics to the following STEM course barriers: 1) challenges with effectively engaging with class content and 2) courses causing severe anxiety.

### A. Challenges effectively engaging with class content

Seven of the nine participants identified practices, or lack thereof, in STEM courses which introduced barriers with effectively engaging with class content. How these barriers arose or interacted with participants' diagnosis characteristics varied but included not having sufficient time to prepare for class, difficulties monitoring learning due to lab sections covering content ahead of the lecture, and course material not seeming relevant to participants' interests.

All three participants diagnosed with ADHD identified requiring more time than their peers to study or complete assignments due to diagnosis characteristics, such as being prone to distractions. For example, participant 1 says: "*Cause generally what takes another person an hour to study it'll take me like 2-4 hours to study… it takes me a while because I just get distracted…*" These participants identified STEM courses as requiring even more time, and that not knowing how to prepare could result in them not having the time necessary to stay on pace with the course. Participant 2 shares how studying takes a particularly long time for physics courses: "*… but with Physics … I spend so much time on the chapter before I do the problem 'cause I really have to get why it's.. you know…why they do certain steps in the example problems and that takes me a long time…*" Participant 2 also described that the SCALE-UP format left them unsure how to prepare: "*I've never been in a class like this [SCALE-UP], so it was kind of different … I wish I would've known how to prepare for the class before 'cause I'm like showing up to the class like any other class...*" Participant 2 shared how these barriers resulted in them being unable to keep pace with the class and subsequently performing poorly on their first exam, "*...and so the test was for um... chapters one through four, but I had only gotten up to chapter one. So, I did get a low grade on a test, but I knew*

---

[2] I (W.J.) acknowledge that my position as lead researcher along with various aspects of my identity (white, male, non-disabled) resulted in me having increased power and persuasion among the researchers involved in analysis.

*everything of chapter one and I knew like half...on chapter two.*" Previous work has also found that students with disabilities report needing increased time for tasks compared to their peers [3,14] and the increased stress this can cause [15]. The participants in this study reveal that STEM courses introduce barriers when this need for processing time interacts with the increased time needed for conceptual understanding. Instructors should account for this increased time when developing and planning curriculum, along with supporting students in developing skills helpful for studying STEM material (e.g. metacognition [16], critical thinking [17]).

Participant 6 experienced challenges with identifying and solving problems in their learning process as a diagnosis characteristic: "*I mean that's an executive functioning thing right there but, I mean just recognizing that there's a problem to even being able to start the problem solving process and to also identify how big the problem is and what resources I need to actually fix it.*" This participant identified executive function as a diagnosis characteristic; executive function is an umbrella term for cognitive functions, such as self-control, working memory, and cognitive adaptability [18]. Participant 6 shared that course structures could make it more difficult to identify and address learning challenges. For example, they shared how the lab section, which occurred separate from the lecture, covered content ahead of the lecture, resulting in the participant not being able to determine if they were on pace with the course: "*because the- the lab and lecture were so disjointed, it always felt like I was behind, even though I wasn't. And... well, I mean, even if I was... I don't know whether I was or not. I really don't.*" The strategies Participant 6 typically used to assess whether they were on pace with a course did not translate to this STEM course; the challenge was significant enough that Participant 6 withdrew from the STEM course. STEM curricula have not typically considered variation in students' executive function [19,20], which creates inequitable learning experiences, such as the one experienced by Participant 6.

When Participant 8 was asked if their disability has shaped their course experience, the participant shared how their chemistry course didn't make connections to their personal interests, which resulted in this participant not engaging with course content: "*I think that it's [their disability] part of why I don't dedicate as much time to it [chemistry course]. Like, it's like if something is what interests me, um what I want to focus on studying, I will put more time towards that 'cause it's more fun to me. And then everything else just gets kicked to the wayside. (laughs).*" Participant 8 expressed how connection to their personal interests could result in high engagement and enjoyment with classes that made the content seem relevant: "*I think that the particular disability helped me kinda thrive was in my environmental science class, because it like kind of allowed me to um... kind of create this hyper-fixation on the subject. Because it was like, it became a very like, a special interest of mine. Like it was easier to keep up with... there was just something about the way that that class went for me was really good....*" Individuals with autism have self-reported a very high engagement with topics of interest [21], and previous studies have found that there is a strong interaction between interest and learning [22]. Our findings reveal the negative and positive consequences this interaction can have in STEM courses for students who identify relevancy as a critical component of their capacity to engage with content.

Our participants described barriers to engaging with the course content arising from not knowing how to prepare for a physics class, misaligned lectures and labs, and content not seeming relevant to personal interests. Though these are likely barriers for many students, our participants experienced severe consequences due to the interaction between the barriers and their diagnosis characteristics. Consequences included being unprepared for assessments and withdrawing from the course. Anxiety was another consequence from barriers to effectively engaging with course content and will be discussed more in the following section.

### B. Courses causing severe anxiety

Four participants identified frequent episodes of anxiety as a diagnosis characteristic and reported that STEM courses increased the frequency and intensity of these episodes. Another three participants experienced episodes of anxiety due to difficulties in preparing for assessments caused by course barriers discussed in the previous section or due to having insufficient time to complete assessments. In both cases, testing accommodations were highly effective in reducing anxiety and were therefore highly valued by participants.

Two of the participants who identified with an anxiety-related diagnosis reported that they received their diagnosis due to experiencing severe episodes of anxiety in previous chemistry courses. Participant 9 shared their experience of a debilitating anxiety episode that occurred before a chemistry exam: "*Yeah. Before one [chemistry] exam, I had such, like I had never ... I've had like anxiety attacks, but I usually calm myself down. But that was the first time that I got an anxiety attack so bad that I just had to go to bed.... Like, I just, I was crying, bawling. And a ton of anxiety, and I just had to go to bed.*" Both participants identified that the anxiety attacks would occur leading up to exams, but only participant 9 identified feeling unprepared for exams as the cause of the episodes of anxiety. Participant 9 expressed course barriers to being prepared for the tests, such as ineffective course resources: "*And plus the answers weren't, when I was doing the practice problems, the answers weren't like explained, So I didn't know what to do. And it was going to take me a ton of time.*" The American College Health Association has found that over 26% of undergraduate students report that their academic performance is negatively affected by anxiety [23]. Our participants reveal that course practices were the instigators of their anxiety, and that this anxiety had severe negative impacts on their emotional health and ability to prepare for assessments.

Test anxiety was also a common experience for participants who didn't identify anxiety as a diagnosis characteristic and could be caused by not having time to process information, seeing others finish before them, or being distracted by noises in the classroom. Eight out of nine participants made use of test accommodations, and seven out of eight of these individuals reported that test accommodations played a significant role in reducing test anxiety. Test accommodations can include having extra test time, having a reduced distraction environment for tests, or being able to make up tests if anxiety attacks prevent their attendance. Regarding the effectiveness of testing accommodations, Participant 1 expressed how having the extra test time dramatically reduced their anxiety by affording them time to not have to rush through the test: "*...knowing I have that extra time makes me, makes my anxiety go down... I feel a lot more comfortable with my answers because I had time to read it, and I wasn't rushing or, you know, it wasn't building up like I'm not going to have enough time...*". Though some participants used testing accommodations for every class, others only used them in courses they identified as challenging. STEM courses often fit this criterion. When asked if they would continue to use accommodations, Participant 9 stated, "*For sure. I'm like, I know like chemistry is a struggle for me. And especially it gets harder if I get anxious. So, for sure. I'm using it for orgo [organic chemistry]. And for math.*" Thus, Participant 9 highlighted STEM courses as a site where the interaction between difficulty and anxiety (a diagnosis characteristic) led to their use of test accommodations. Previous studies have found that accommodations are highly important for students with disabilities, but instructors can present barriers to their use by expressing negative accommodation viewpoints. such as accommodations being unnecessary [3,15,24]. The participants in this study report accommodations as often being more important in STEM courses, and so it is critical that STEM instructors welcome and encourage their use. Instructors can make students feel more comfortable accessing accommodations by inviting students to discuss their needs on the first day while discussing the syllabus or a few weeks before each assessment (as there is often a time lag between requesting and gaining access to accommodations).

Anxiety was commonly reported among our participants, most often experienced before or during STEM tests. The experienced episodes of anxiety varied in intensity, but always resulted in negative consequences including emotional distress or performing worse on assessments. Participants identified that testing accommodations were critical to the reduction of anxiety.

## IV. IMPLICATIONS

These findings reveal that there are significant existing barriers in STEM courses for students with disabilities. By centering students who experience these barriers, we can begin identifying and addressing the barriers. While we identified barriers in relation to participants' diagnosis characteristics, it is likely that these challenges also exist for students without disability diagnoses. By implementing course practices that support students with disabilities, we provide more support for all students; this idea is a driving motivation of the Universal Design for Learning (UDL) framework [8]. By using recommendations from the UDL framework, we can identify strategies and practices to address course barriers [25].

To support students in effectively studying, the UDL framework suggests highlighting critical features and big ideas and providing graphic organizers or checklists or guides for notetaking. By highlighting what is important for students to learn, instructors help students to spend their time effectively on content that is relevant to the learning objectives. Providing graphic organizers or note guides can help scaffold the development of the skills necessary to learn STEM content, while also reducing the load on executive functions by providing tools that help students monitor their learning. Our participants suggested that instructors provide weekly quizzes, structured similarly to the tests, to provide feedback on students' progress. For example, Participant 9 says, "*Because I prefer having quizzes weekly. Where I know I'm testing my knowledge...*"

UDL-aligned strategies to support students in coping with stress and anxiety include promoting external emotional support (e.g., campus counseling services) or shifting questions of natural aptitude to questions of how a student can improve. Negative stigmas concerning counseling services and mental health are often the greatest barrier to students pursuing help [26], so instructors should frame mental health support as natural rather than tied to individual deficit [27]. Many college students do not know about university provided mental health services [28], so instructors should present and remind students of campus resources. Participants suggested that instructors could help to reduce their anxiety by providing study guides or practice tests, which, like weekly quizzes, help students monitor their progress. Participant 4 expresses this saying, "*Like having ways to actually apply the material, to have practiced things to let you know if you know it, that's so helpful, so helpful.*"

A limitation of the UDL framework is that it is not content specific. Thus, researchers should continue to investigate the content specific experiences of students with disabilities to guide development, implementation and evaluation of disciplinarily-specific UDL-aligned practices.

## V. ACKNOWLEDGEMENTS

This work is supported by NSF DUE No. 1612009.